# Symmetry regimes for circular photocurrents in monolayer MoSe$_2$

Jorge Quereda[1]*, Talieh S. Ghiasi[1], Jhih-Shih You[2], Jeroen van den Brink[2], Bart J. van Wees[1], Caspar H. van der Wal[1]

[1] Zernike institute for Advanced Materials, University of Groningen, NL-9747AG Groningen, The Netherlands
[2] Institute for Theoretical Solid State Physics, IFW Dresden, Helmholtzstr. 20, 01069 Dresden, Germany

**ABSTRACT**

**In monolayer transition metal dichalcogenides helicity-dependent charge and spin photocurrents can emerge, even without applying any electrical bias, due to circular photogalvanic and photon drag effects. Exploiting such circular photocurrents (CPC) in devices, however, requires better understanding of their behavior and physical origin. Here, we present symmetry, spectral, and electrical characteristics of CPC from excitonic interband transitions in a MoSe$_2$ monolayer. The dependence on bias and gate voltages reveals two different CPC contributions, dominant at different voltages and with different dependence on illumination wavelength and incidence angles. We theoretically analyze symmetry requirements for effects that can yield CPC and compare these with the observed angular dependence and symmetries that occur for our device geometry. This reveals that the observed CPC effects require a reduced device symmetry, and that effects due to Berry curvature of the electronic states do not give a significant contribution.**

*Introduction*

Among two-dimensional (2D) materials, monolayer transition metal dichalcogenides (1L-TMDCs) offer a versatile platform for the development of spintronic and valleytronic devices, where the spin and valley degrees of freedom are used as information carriers.[1–5] The particular band structure of 1L-TMDCs, where two nonequivalent valleys appear at the K and K' points of the 2D Brillouin zone, gives rise to valley-dependent optical selection rules. Specifically, when a 1L-TMDC is illuminated with circularly polarized light with a photon energy close to its bandgap, optical transitions can only take place in one of the two valleys, either K or K', depending on the helicity of the circular polarization, leading to a light-induced valley population imbalance.[5] Additionally, monolayer TMDCs present a large spin-orbit splitting, which sign changes between the K and K' valleys, causing a coupling between the spin and valley degrees of freedom.[5] As a consequence, different optical processes can be used to generate spin and valley polarized photoresponse in TMDCs, such as the valley-Hall effect [6,7]. For this effect, under circularly polarized illumination, charge carriers in different valleys flow to opposite transverse edges when an in-plane electric field is applied, producing a light helicity-dependent Hall voltage.

The recent observation of helicity-sensitive circular photogalvanic effect (CPGE) [8,9], both for multilayer and monolayer TMDCs [10,11], opens another route for producing spin-valley transport through a 1L-TMDC phototransistor. Differently from the valley-Hall effect, which relies on applying an in-plane voltage gradient to the TMDC in order to obtain spin-valley current, the CPGE allows to generate a directed spin-valley current even without applying any voltage, bringing new opportunities for the implementation of active spintronic and valleytronic devices. However, a comprehensive study of this effect and its microscopic origin in 1L-TMDCs is still missing.

In this work we investigate for the first time the spectral and electrical behavior of the helicity-dependent circular photocurrent (CPC) in a 1L-TMDC. By evaluating the spectral response of the CPC in a h-BN encapsulated 1L-MoSe$_2$ phototransistor, we show that the CPC amplitude is maximized when the illumination wavelength matches the A exciton resonance, clearly demonstrating that excitonic absorption plays a central role in the generation of the CPC. We also explore the effect of a drain-source voltage ($V_{ds}$) on the CPC, revealing two different regimes for small (below 0.4 V) and large voltages, with the CPC changing sign between one regime and the other. Further, we find that the CPC presents very different symmetry upon change of the light incidence angle for the two regimes: For small $V_{ds}$, the CPC is preserved when the incidence angle is switched from $\phi$ to $-\phi$, whereas for large $V_{ds}$, inverting the illumination angle $\phi$ causes a change of sign for the CPC, pointing to a separate physical origin. Recently, it was described[11] that Berry-curvature (BC) at the band edges of 1L-TMDCs can give a contribution to CPC (BC-induced circular photogalvanic effect, BC-CPGE). However, we find that BC-CPGE is not compatible with the angular dependences observed for any of the two CPC regimes. Further, we show that CPC can also emerge in this system due to the circular photon drag effect (CPDE), mostly overlooked in prior literature for 1L-TMDCs. Finally, we show how by applying a gate voltage to modify the Fermi energy of the 1L-MoSe$_2$ channel, one can tune the relative strength of the two contributions at a fixed drain-source voltage, achieving control over the intensity and direction of the helicity-dependent photoresponse.

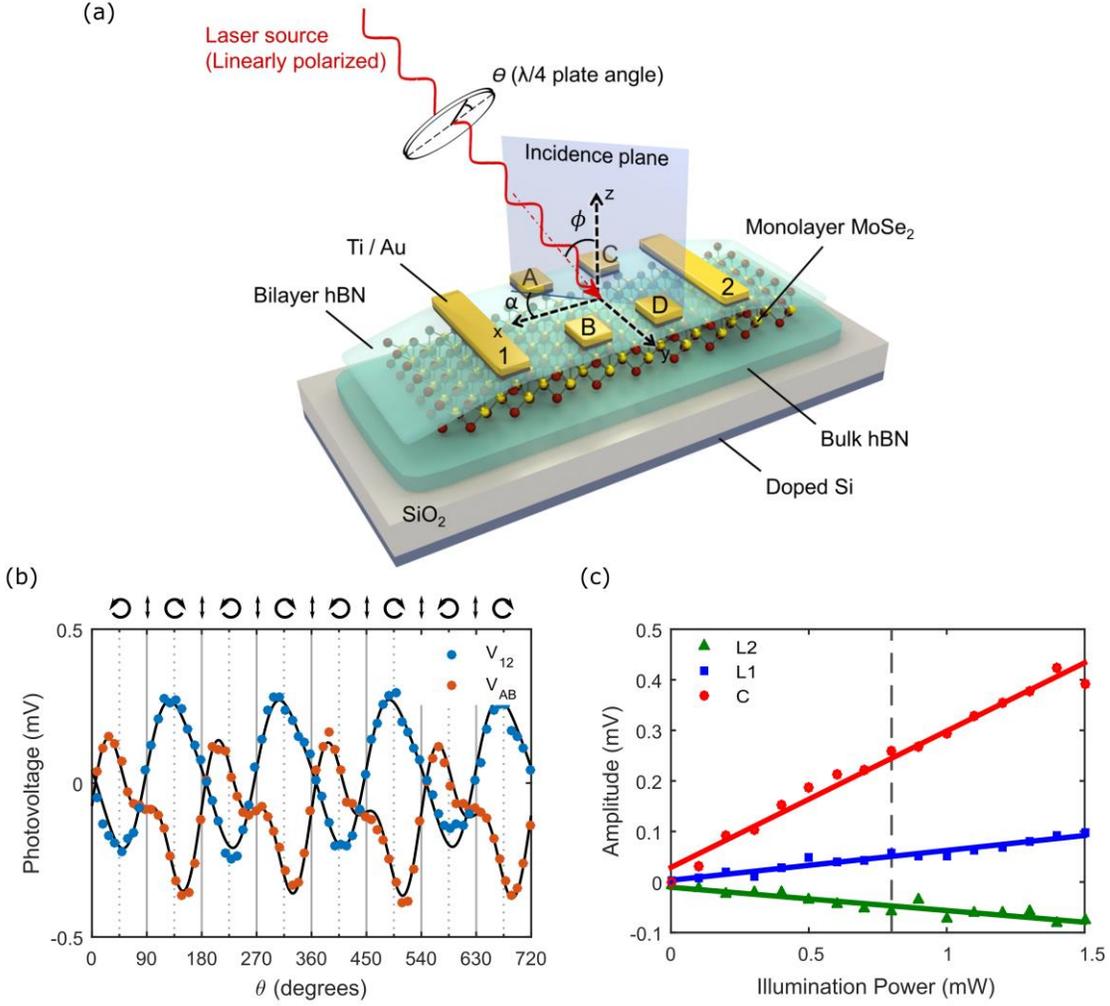

**Fig 1 –** (a) Schematic experimental setup. The helicity of the laser excitation is controlled by rotating the quarter-wave plate angle, $\theta$. (b) Helicity-dependent photovoltage of the contacts [1, 2] (blue) and [A, B] (orange) as a function of the quarter-waveplate angle $\theta$ for $\lambda$ = 785 nm, $\phi$ = 20°, $V_{ds}$ = 0, $V_{gate}$ = 0 and $\alpha$ = 45°. The black lines are fits to the phenomenological equation (1). (c) Power dependence of $C$, $L_1$ and $L_2$ (extracted from fits to equation (1)). The solid lines are linear fits to the experimental data. The vertical dashed line indicates the power used during the experiments, 0.8 mW.

*Circular photocurrent in 1L-TMDC for interband transitions.*

When spatial inversion symmetry is broken in the 2D plane in a system with time-reversal symmetry, illumination with circular light can generate a DC photocurrent $\vec{J}$ that behaves as a second order response to the electric field. $\vec{J}$ can be written as a series expansion in the light wave vector $\vec{q}$ as $J_l = \chi_{ljk} E_j E_k^* + T_{ljk\mu} q_\mu E_j E_k^* + (...)$. Here $\chi_{ljk}$ and $T_{ljk}$ are the photogalvanic and photon drag susceptibility tensors and $l, j, k, \mu$ label Cartesian coordinates $x, y$ and $z$. As we present in Suppl. Info. S6, the device symmetries strongly constrain the tensor components, and they can still vanish for high-symmetry configurations, even for broken inversion symmetry. We consider three different symmetry scenarios: $D_{3h}$ (pristine 1L-MoSe$_2$), $C_{3v}$ (1L-MoSe$_2$ with broken out-of-plane mirror symmetry), and single-mirror symmetry (1L-TMDCs in presence of strain or device inhomogeneities). Comparing the dependence of CPC on illumination angles with the symmetry-allowed CPGE and CPDE contributions, we find that for the low-bias regime our observations are only compatible with a device symmetry of, at most, a single mirror plane. For the high-bias regime the CPC effects are also compatible with $C_{3v}$ symmetry.

In previous reports, the CPC in 1L-TMDCs has been associated with a BC-induced CPGE.[11,12] In 1L-TMDCs, the BC takes opposite signs at the K and K' valleys, giving rise to counterpropagating valley currents.[5,6,13] Thus, when circularly polarized illumination is used to produce a valley population imbalance, a CPC contribution can appear. In Suppl. Info. S6 we derive the CPGE photocurrent using the Fermi Golden rule. This shows that resonant interband transitions can produce a BC contribution to the CPGE [14]. However, this contribution should maximize for incidence perpendicular to the 2D plane[12], while our experiments only show nonzero CPC at oblique incidence (see below).

*Device fabrication, electrical characterization and setup*

The fabricated 1L-MoSe$_2$ field-effect transistor is depicted in **Fig. 1a** and the actual device is shown in the Suppl. Info. (Figure S2). To improve the device quality and stability [15,16], the 1L-MoSe$_2$ channel is encapsulated between a bilayer and a bulk h-BN flake, acting as the top and bottom layers respectively. The 2L-BN/1L-MoSe$_2$/bulk-BN stack is directly built onto a SiO2/doped Si substrate with an oxide thickness of 300 nm. The electrodes are fabricated on top of the structure by e-beam lithography (EBL) and e-beam evaporation of Ti (5 nm)/Au (55 nm) (see Methods section). To further avoid the presence of adsorbates and contaminants, the sample is kept in vacuum (10$^{-4}$ mbar) during the whole experiment. All experiments were carried out at room temperature.

Electrical characterization of the sample (Suppl. Info. S1) identifies n-type character for the 1L-MoSe$_2$, with a threshold gate voltage of about 20 V and an electron mobility of 17 cm²/V.s. In this sample geometry, the bilayer h-BN plays the role of a tunnel barrier, preventing Fermi level pinning at the metal-semiconductor interfaces.[17]

*Helicity-resolved photovoltage measurements: description and phenomenological formula.*

**Figure 1a** depicts the experimental setup for measuring the helicity-dependent photogalvanic response of the MoSe$_2$ phototransistor. We illuminate the sample at an oblique angle $\phi$ with respect to the normal vector of the crystal surface and simultaneously measure the photoinduced currents, either directly (Suppl. Info. S5) or as the associated voltages (main text). We used two perpendicular sets of electrodes, giving voltage signals $V_{12}$ and $V_{AB}$. For illumination, we used a laser with tunable photon energy. For achieving a uniform illumination power density and well-defined light incidence angles, we used a collimated beam of 0.5 cm diameter, much larger than the studied device. The polarization of the illumination beam was tuned by rotating a $\lambda/4$ waveplate over an angle $\theta$: during rotation over 360° the original linear polarization gets modulated twice between left and right circular polarization (see top labels Fig. 1b).

**Figure 1b** shows $V_{12}$ and $V_{AB}$ as a function of $\theta$ for illumination fixed at 785 nm (1.58 eV, on-resonance with the $A^0$ exciton transition of monolayer MoSe$_2$ [2,18,19]), incidence angle $\phi = 20°$ and azimuthal angle $\alpha = -45°$ (defined as the angle between the x axis and the incidence plane, see Fig. 1a). The gate voltage was fixed to $V_{gate} = 0$ V. Both voltages clearly show a polarization dependence, with $2\theta$ and $4\theta$-periodic components. The fingerprint of a CPC contribution is its helicity-dependence, appearing as a signal $V_{CPC} \propto \sin(2\theta)$. A $4\theta$-periodic modulation, $V_{LPC}$, can also appear due to linear photogalvanic and linear photon drag effects[8]. The total photovoltage $V_{PC}$ can be described phenomenologically as[8,11,20,21]

$$V_{PC} = V_0 + C\,\sin(2\theta) + L_1 \sin(4\theta) + L_2 \cos(4\theta), \qquad (1)$$

where $C$ accounts for the CPC and $L_1$ and $L_2$ account for the linear photogalvanic and photon drag effects. The total linear polarization-dependent contribution can be accounted as $L = (L_1^2 + L_2^2)^{1/2}$. An additional polarization-independent term, $V_0$, (typically smaller or, at most, comparable to $C$) can also appear due to inhomogeneities or thermal drifts between the two electrodes. We obtain values for $C$, $L_1$ and $L_2$ by fitting equation (1) to data as in Fig. 1b. **Figure 1c** shows the power dependence of $C$, $L_1$ and $L_2$. The three amplitudes increase linearly with the illumination power, confirming that they are due to a second order response to the light electric field, and in agreement with earlier literature for 1L-MoS$_2$.[11]

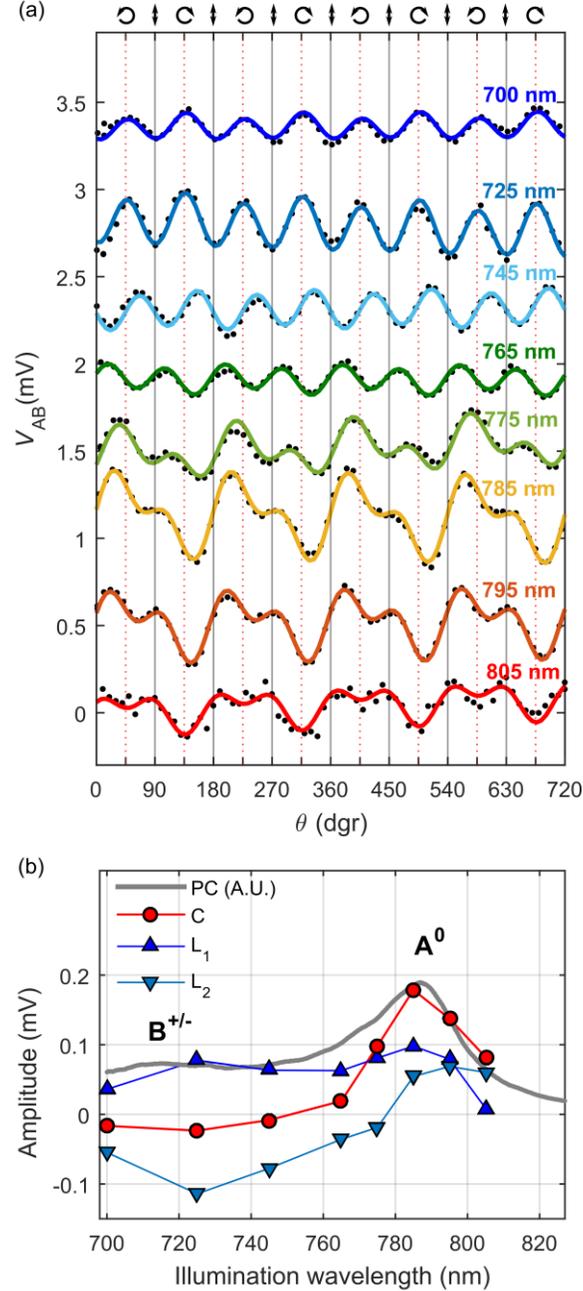

**Fig 2** – Spectral evolution of the circular photocurrent. (a) $V_{AB}$ as a function of the waveplate angle, $\theta$ (for $\phi = 20°$, $V_{ds,12} = 0$, $V_{gate} = 0$ and $\alpha = 45°$) under different illumination wavelengths, from 700 nm to 825 nm. For clarity, the traces have been vertically shifted in steps of 0.5 mV. The solid lines are fits to equation (1). (b) Photocurrent spectrum of the 1L-MoSe$_2$ crystal (grey, solid line) and spectral dependence of the fitting parameters $C$, $L_1$ and $L_2$ (red, dark blue and pale blue lines, see legend).

Finally, we remark that the CPC signal $C$ behaves as reported below for multiple electrode configurations. We can thus rule out that our CPC signals emerge due to properties of specific contacts, or effects from confinement of light between the micron-scale metallic electrodes. We elaborate on this in Suppl. Info. S5.

*Spectral response of the CPC*

Next, we investigate the spectral response of the observed helicity-dependent photovoltage. We first characterize the spectral features of the monolayer MoSe$_2$ phototransistor by photocurrent spectroscopy [2] (see ref. [19] for detailed discussion about our measurement technique). We illuminate the sample using a linearly-polarized continuous-wave tunable infrared laser and register the photovoltage as a function of the illumination wavelength at a constant drain-source bias, $V_{ds}$ = 1 V. The resulting photocurrent spectrum (grey line in **Figure 2b**), shows a prominent peak at 1.58 eV (785 nm), corresponding to the A$^0$ exciton resonance of MoSe$_2$. A second, less prominent, peak can also be observed at 1.74 eV (713 nm), which results from the B$^{+/-}$ trion transition [2,19].

**Figure 2a** shows the helicity-dependent photovoltage of the 1L-MoSe$_2$ device and fits to equation (1) for different illumination wavelengths. **Figure 2b** shows the wavelength dependence of $C$, $L_1$ and $L_2$. The CPC contribution $C$ is maximal when the illumination is on-resonance with the A$^0$ exciton transition ($\lambda$ = 785 nm) and progressively decreases when the illumination is shifted away from the resonance. For the linear photovoltage $L$ a nonzero amplitude appears, also for out-of-resonance illumination.

The observed spectral behavior of $C$ shows that interband excitons play a central role in the CPC photoresponse. However, since excitons are charge-neutral quasiparticles, they must dissociate to produce a nonzero photocurrent. The required dissociation can be assisted by the large in-plane electric fields present in the depletion regions near a metal-semiconductor junction, especially when a bias voltage is applied[7]. Alternatively, a photocurrent can appear in absence of in-plane electric fields if trions are present in the MoSe$_2$, since they have a nonzero net charge, and can contribute to the photocurrent even without dissociating. Since in our system the whole device is illuminated, both dissociated excitons and non-dissociated trions are expected to play a role in the CPC.

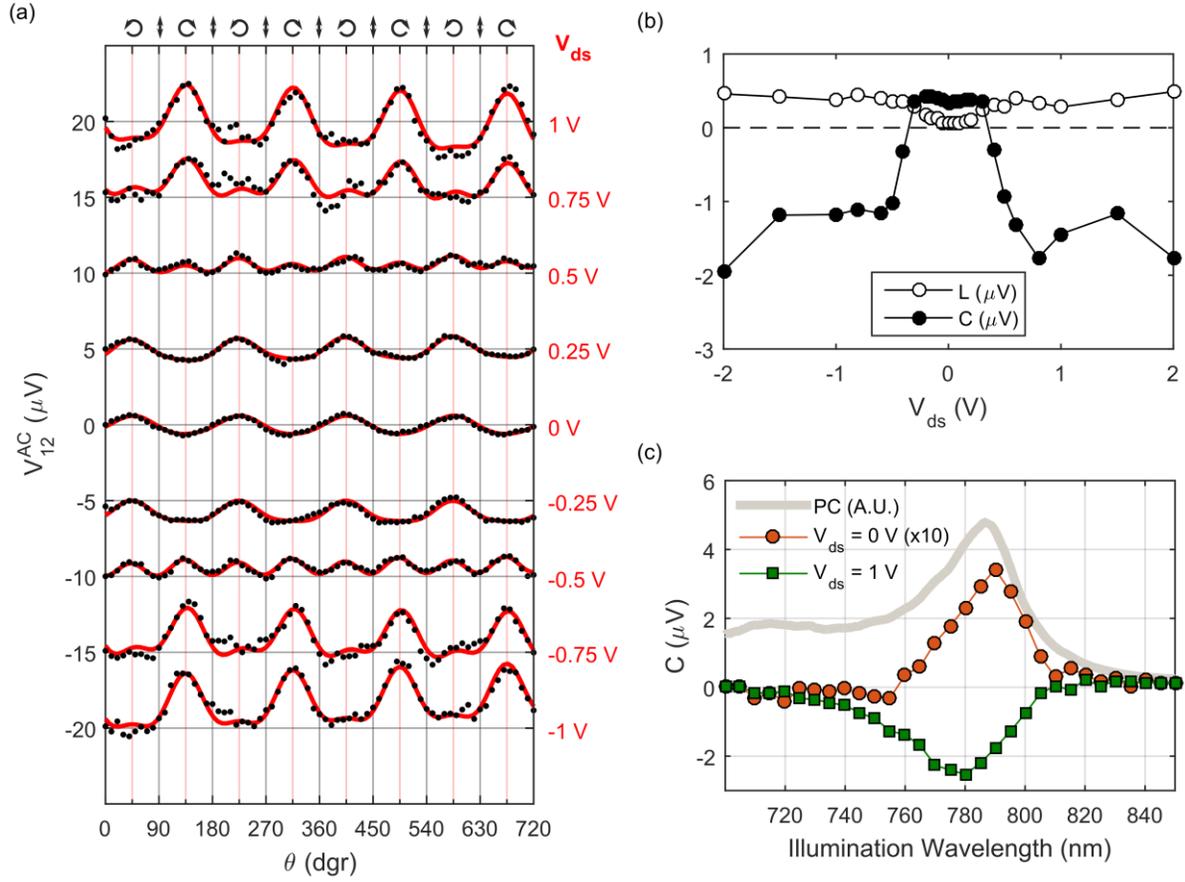

**Fig 3 –** Helicity-dependent photovoltage, $V_{12}^{AC}$ for different drain-source voltages, $V_{ds}$. (a) $V_{12}^{AC}$ as a function of $\theta$ for different drain-source voltages with $\lambda$ = 785 nm, $\phi$ = 20°, $V_{gate}$ = 0 and $\alpha$ = 45°. For clarity, the measurements have been vertically shifted in steps of 5 µV and the polaritazion-independent offset, $V_0$, has been substracted (see equation (1)). (b) $C$ and $L$ parameters as a function of the drain-source voltage. (c) CPC amplitude, $C$, as a function of the wavelength for $V_{ds}$ = 0 V (orange circles) and $V_{ds}$ = 1 V (green squares). For an easier visualization, the data for $V_{ds}$ = 0 V has been multiplied by 10.

*Effect of a nonzero drain-source voltage.*

To investigate the influence of an in-plane electric field on the CPC we apply a drain-source voltage $V_{ds}$ between the electrodes A and B and measure the transverse voltage between the electrodes 1 and 2, while keeping $\lambda$ = 785 nm, $V_{gate}$ = 0, $\phi$ = 20° and $\alpha$ = 45°. For improving the signal-to-noise ratio, we now use a chopper to modulate the laser intensity at 331 Hz and lock-in detection of the AC photovoltage $V_{12}^{AC}$. **Figure 3a,b** show the helicity dependence of $V_{12}^{AC}$ at different $V_{ds}$ and the associated dependence of $C$ and $L$ on $V_{ds}$. Unlike the case of the valley-Hall effect (where the anomalous Hall voltage changes linearly with the applied drain-source voltage) the CPC response observed here preserves its sign when the direction of the drain-source voltage is inverted. For small applied voltages, up to $|V_{ds}|\sim 0.4$ V $\equiv V_T$ (transition voltage), $C$ remains constant. When increasing $|V_{ds}|$ above $V_T$ the photogalvanic signal undergoes an abrupt change of sign and becomes ~5-10 times larger.

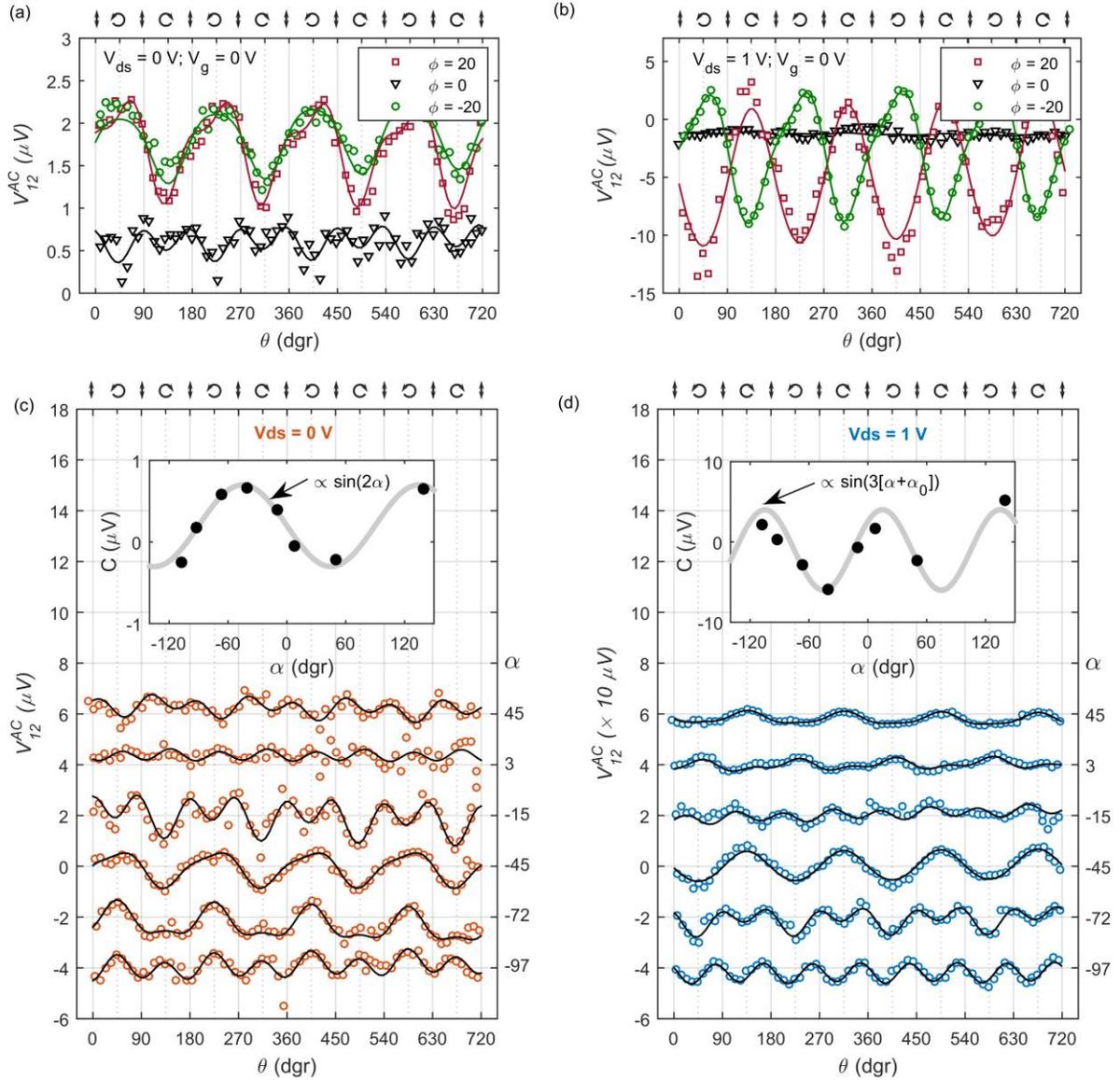

**Fig 4** – Effect of the illumination angle on the CPC amplitude, $C$. (a) Helicity-dependent photovoltage measured at $V_{ds}$ = 0 V, $V_{gate}$ = 0 V and $\lambda$ = 785 nm for three different illumination incidence angles, $\phi$ = -20°, 0° and 20°. The azimuthal angle, $\alpha$, is fixed at $\alpha$ = 45°. (b) Same as (a) for $V_{ds}$ = 1 V. (c) Helicity-dependent photovoltage at different azimuthal angles, $\alpha$, with $V_{ds}$ = 0 V and $\phi$ = 20°. For clarity, the helicity-independent background has been removed and the plots have been shifted vertically in steps of 2 $\mu V$. Inset: CPC amplitude, $C$, extracted by fitting the measured photovoltage to equation (1), as a function of $\alpha$. We observe that $C$ changes proportionally to $\sin(2\alpha)$. (d) Same as (c) for $V_{ds}$ = 1 V. In this case, $C$ changes proportionally to $\sin(3\alpha)$.

**Figure 3c** shows the CPC amplitude $C$ as a function of the wavelength at $V_{ds} = 0$ V and $V_{ds} = 1$ V. Interestingly, the wavelength at which $C$ is maximized (in absolute value) changes from 790 nm (1.57 eV) at $V_{ds} = 0$ V to 780 nm (1.59 eV) at $V_{ds} = 1$ V. This suggests that at low drain-source voltages the dominant charge carriers involved in the CPC are $A^{+/-}$ trions (which have a nonzero charge and therefore do not need to dissociate to participate in the photovoltage), while at large drain-source voltages the transport is dominated by dissociated $A^o$ excitons.[22]

*Effect of the illumination angle in the CPC*

In order to identify the symmetry properties of the two different CPC regimes (for $V_{ds}$ above and below $V_T$) we test their behavior under different illumination angles. **Figure 4** shows the measured helicity-dependent photovoltage $V_{12}^{AC}$ for different illumination incidence angles $\phi$, in the low-$V_{ds}$ (4a) and high-$V_{ds}$ (4b) regimes. Remarkably, these two regimes show a very different behavior: For $V_{ds} = 0$ V, the CPC shows the same sign and a similar amplitude at $\phi = 20°$ and $\phi = -20°$, while, for $V_{ds} = 1$ V, inverting the angle of incidence causes the CPC to reverse its sign, pointing to two separate physical mechanisms. Importantly, for both situations $C$ vanishes for incidence normal to the 2D plane, $\phi = 0°$, which rules out that BC-induced CPC gives significant contributions to our signals (Suppl. Info. S6).

We further check the symmetry of the measured CPC by characterizing its dependence on the azimuthal angle $\alpha$ (see Fig. 1a). **Figures 4c,d** show the measured helicity-dependent photovoltages at different azimuthal angles, for $|V_{ds}|<V_T$ (4c) and $|V_{ds}|>V_T$ (4d). The insets show the dependence of $C$ on $\alpha$. Again, two different behaviors emerge: For small $V_{ds}$, $C$ is proportional to $\sin(2\alpha)$. We remark that, since the CPC sign is preserved upon inversion of $\phi$, it must also be preserved upon a $\pi$ rotation of $\alpha$ (both operations are equivalent in our system), and therefore, only a $\pi$-periodic dependence on $\alpha$ can appear.

For large $V_{ds}$, $C$ shows a modulation proportional to $\sin(3[\alpha+ \alpha_o])$, where $\alpha_o$ is an angle offset (15° in our case). This $3\alpha$-periodic signal suggests that $C$ is modulated by the 120°-periodic crystal structure of 1L-MoSe$_2$. The presence of an angular offset $\alpha_o$ is expected since the orientation of the crystal is not necessarily aligned with the electrodes. Oppositely from before, only an $\alpha$-dependence that gives an exact inversion upon $\pi$ increase of $\alpha$ can emerge, for consistency with signal inversion when $\phi$ is reversed.

As discussed in the Suppl. Info S6, when the device symmetry is reduced to, at most, a single-mirror symmetry (which can be expected in a realistic device due to interface effects at the electrodes and in-plane strain gradients), a CPDE photocurrent can have a term proportional to $\sin(2\alpha)\sin^2(\phi)$, consistent with the observed behavior at low $V_{ds}$. For the large $V_{ds}$ regime, the inversion of the CPC upon sign flip of $\phi$ is consistent with both CPGE and CPDE terms (or a combination of them) allowed for this symmetry. Further, $\phi$-odd terms are also allowed for CPGE and CPDE under the more restrictive $C_{3v}$ symmetry. Notably, the dependence as $\sin(3\alpha)$ for the CPC measured at large $V_{ds}$ does not appear in the symmetry analysis. Such dependence, however, can emerge from inhomogeneities of the transport properties between the armchair and zigzag directions of the 1L-MoSe$_2$ crystal lattice, not considered in the theory.

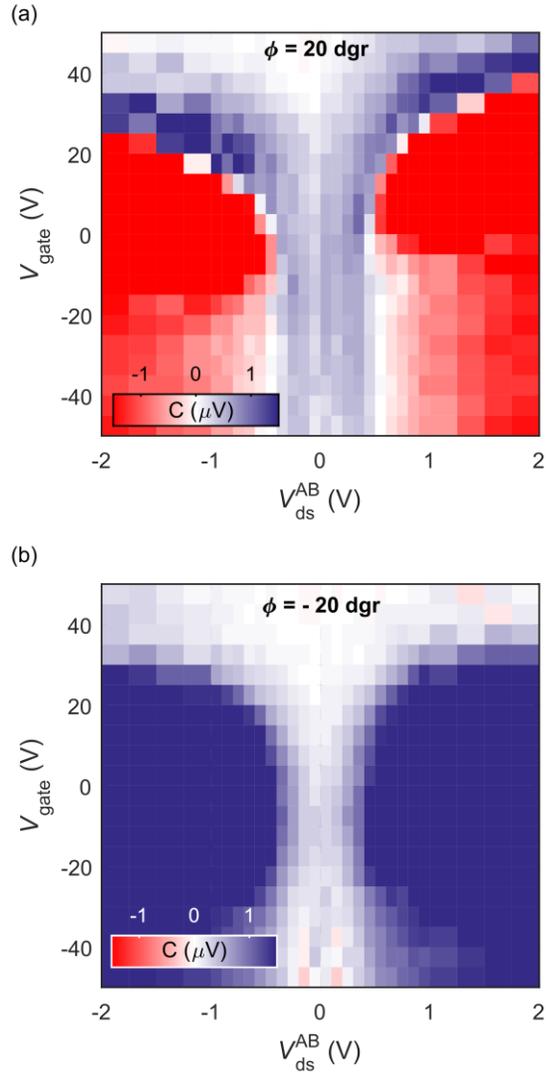

**Fig 5** – (a) Colormap of the CPC amplitude, $C$ for $\lambda$ = 785 nm as a function of the drain-source and gate voltages, $V_{ds}$ and $V_{gate}$ for $\phi$ = 20°. (b) Same as (a) for an incidence angle $\phi$ = - 20°.

*Effect of the gate voltage in the CPC*

Finally, we explore how the CPC is affected by the gate voltage. **Figure 5a** shows a color map of the CPC amplitude $C$ (derived from $V_{12}^{AC}$ lock-in signal) as a function of $V_{ds}$ (applied between electrodes A and B) and $V_{gate}$, at $\alpha$ = - 45° and $\phi$ = 20°. The two drain-source voltage regimes discussed above can be observed here as the blue ($C > 0$ mV) and red ($C < 0$ mV) areas of the map. Once again, when the incidence angle $\phi$ is changed to -20° (see Figure 5b), the sign of $C$ at large $V_{ds}$ switches from negative to positive, while at small $V_{ds}$ the sign is preserved.

For $V_{gate}$ below 0 V we see a much weaker influence on the CPC than for $V_{gate}$ > 0 V. In the latter case we observe a shift of the transition voltage $V_T$ towards larger drain-source voltages. This can be explained by an increased trion population, due a higher density of charge carriers in the MoSe$_2$ crystal when the Fermi energy is brought above the edge of the conduction band [22]. Further, an increased gate voltage can also modify the electric field screening, changing the exciton and trion momentum lifetimes and therefore changing their contributions to the CPC.[11,12] When the gate voltage is further increased we observe an overall reduction of the CPC, regardless of the value of $V_{ds}$, which we associate to a decrease of the carrier momentum lifetime, due to an enhanced electron-electron scattering. Also, the probability of exciton absorption is expected to decrease at large gate voltages, due to the reduced density of unoccupied states in the conduction band.

*Summary and conclusions*

In conclusion, the two observed regimes for the CPC can be well-described by CPGE and CPDE for a reduced device symmetry. Although effects of higher order in the light electric field could also be allowed by symmetry, the linearity of $C$ with illumination power confirms that the measured signal is dominated by second-order effects.

Importantly, although a Berry-curvature CPGE could be allowed for a low-symmetry device, it is not observed here, as confirmed by the fading of $C$ for incidence normal to the crystal plane. Further, our results indicate a transition from exciton- to trion-dominated transport between the two regimes, but the influence of the excitonic character on CPC is an open question.

## METHODS

*Device Fabrication*

We mechanically exfoliate atomically thin layers of MoSe$_2$ and h-BN from their bulk crystals on a SiO$_2$ (300 nm)/doped Si substrate. The monolayer MoSe$_2$ and bilayer h-BN are identified by their optical contrasts with respect to the substrate [23] and their thickness is confirmed by by Atomic Force Microscopy (see SI, Figure S2). Using a polymer-based dry pick-up technique, described in detail in ref. [24], we pick up the bilayer h-BN flake using a PC (Poly(Bisphenol A)carbonate) layer attached to a PDMS stamp. Then we use the same stamp to pick up the MoSe$_2$ flake directly in contact with the h-BN surface and we transfer the whole stack onto a bulk h-BN crystal, exfoliated on a different SiO$_2$/Si substrate. After the final transfer step, the PC layer is detached from the PDMS, remaining on top of the 2L-BN / MoSe$_2$ / bulk-BN stack, and must be dissolved using chloroform. Next, to further clean the stack, we anneal the sample in Ar/H$_2$ at 350 °C for 3 hrs. For the fabrication of electrodes, we pattern them by electron-beam lithography using PMMA as the e-beam resist, followed by e-beam evaporation of Ti(5 nm)/Au(75nm) at 10$^{-6}$ mbar and lift-off in Acetone at 40°C.

*Electrical characterization*

The DC electrical characterization of the studied device is discussed in detail in section S1 of the Supporting Information. The highly-doped Si substrate is used as the back-gate electrode in order to tune the density of charge carriers in the MoSe$_2$ channel. To eliminate the effect of environmental adsorbates, all the electrical measurements are performed in vacuum (~10$^{-4}$ mbar). We measure the source-drain current as a function of the source-drain and back-gate voltages in four-terminal geometry of electrodes, using the side contacts of the Hall-bars[25] as voltage probes. These measurements allow us to obtain a reliable estimation of conductivity and field effect mobility of charge carriers in the monolayer MoSe$_2$ channel. Further I-V characteristics are measured in 3-terminal geometry to evaluate quality the electrical contacts at the metal-semiconductor interface, as further discussed in section S1.

## ASSOCIATED CONTENT

The following associated content can be found in the supporting information to this article:

S1. AFM characterization

S2. Optical microscopy images of the fabrication process and the final device

S3. Electrical characterization of the 1L-MoSe$_2$ phototransistors

S4. Color map of the CPC amplitude as a function of $V_{ds}$ and $V_{gate}$ for illumination at normal incidence

S5. Comparison of the photovoltage and photocurrent measurements and consistency checks

S6. Theoretical analysis of the photogalvanic and photon drag effects

## AUTHOR INFORMATION

### Corresponding Author

* j.quereda.bernabeu@rug.nl

### Author Contributions

BJW and CHW initiated the project. JQ and TSG had the lead in experimental work. JSY and JB provided the theoretical analysis. The manuscript was written and reviewed by all authors.

### Funding Sources

This research has received funding from the Dutch Foundation for Fundamental Research on Matter (FOM), as part of the Netherlands Organisation for Scientific Research (NWO), FLAG-ERA (15FLAG01-2), and the Zernike Institute for Advanced Materials.

## ACKNOWLEDGMENTS

We thank Feitze A. van Zwol, Tom Bosma and Jakko de Jong for contributions to the laser control system. We thank H. M. de Roosz, J. G. Holstein, H. Adema and T. J. Schouten for technical assistance.

## ABBREVIATIONS

TMDC, Transition metal dichalcogenide; CPGE, Circular Photogalvanic Effect; LPGE, Linear Photogalvanic Effect; LPDE, Linear Photon Drag Effect; CPC, Circular Photocurrent, BC; Berry curvature.

# Supplementary information to:
# Symmetry regimes for circular photocurrents in monolayer MoSe$_2$

Jorge Quereda[1*], Talieh S. Ghiasi[1], Jhih-Shih You[2], Jeroen van den Brink[2], Bart J. van Wees[1], Caspar H. van der Wal[1]

[1] Zernike institute for Advanced Materials, University of Groningen, NL-9747AG Groningen, The Netherlands
[2] Institute for Theoretical Solid State Physics, IFW Dresden, Helmholtzstr. 20, 01069 Dresden, Germany

**CONTENT**

## S1. AFM characterization

We measure the height profile of the BN-encapsulated MoSe$_2$ on a SiO$_2$/Si substrate by AFM. The thickness of both of the MoSe$_2$ and the top h-BN flakes are measured as 0.7 nm (Figure S1b) which corresponds to monolayer MoSe$_2$ and bilayer h-BN, in agreement with the reported values in literature[1,2]. The bottom h-BN has a thickness of 7.65 nm (21-22 layers). The AFM images also reveal the presence of bubbles due to trapped molecules in the h-BN/MoSe$_2$ interface. Reportedly, the accumulation of the interface contaminants in these bubbles ensures a perfectly clean interface at the bubble-free regions, and is a signature of the good adhesion between the two layers.[3]

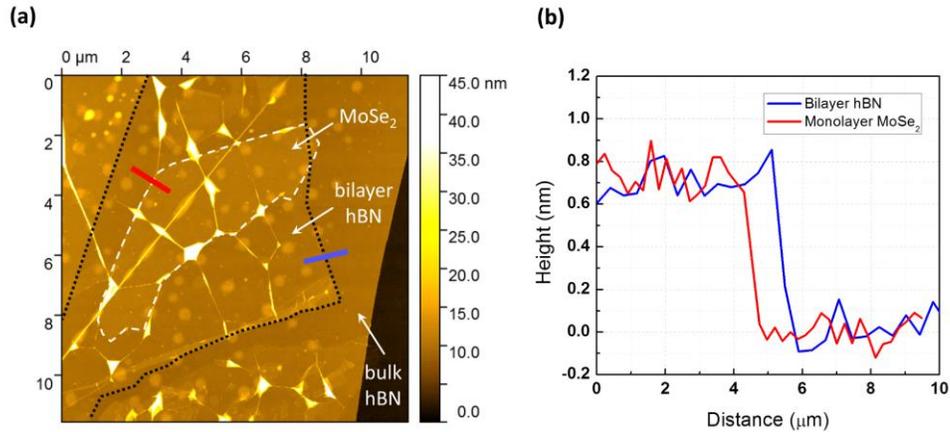

**Fig S1** (a) AFM image of the BN-encapsulated MoSe$_2$ on SiO$_2$/Si substrate. The dashed lines highlight the edge of the flakes. (b) Height profile along the red and blue lines indicated in panel (a), corresponding to the edges of the monolayer MoSe$_2$ and the bilayer h-BN flakes. For clarity, The profiles are offsetted to the same zero level.

## S2. Optical microscopy images of the fabrication process and the final device

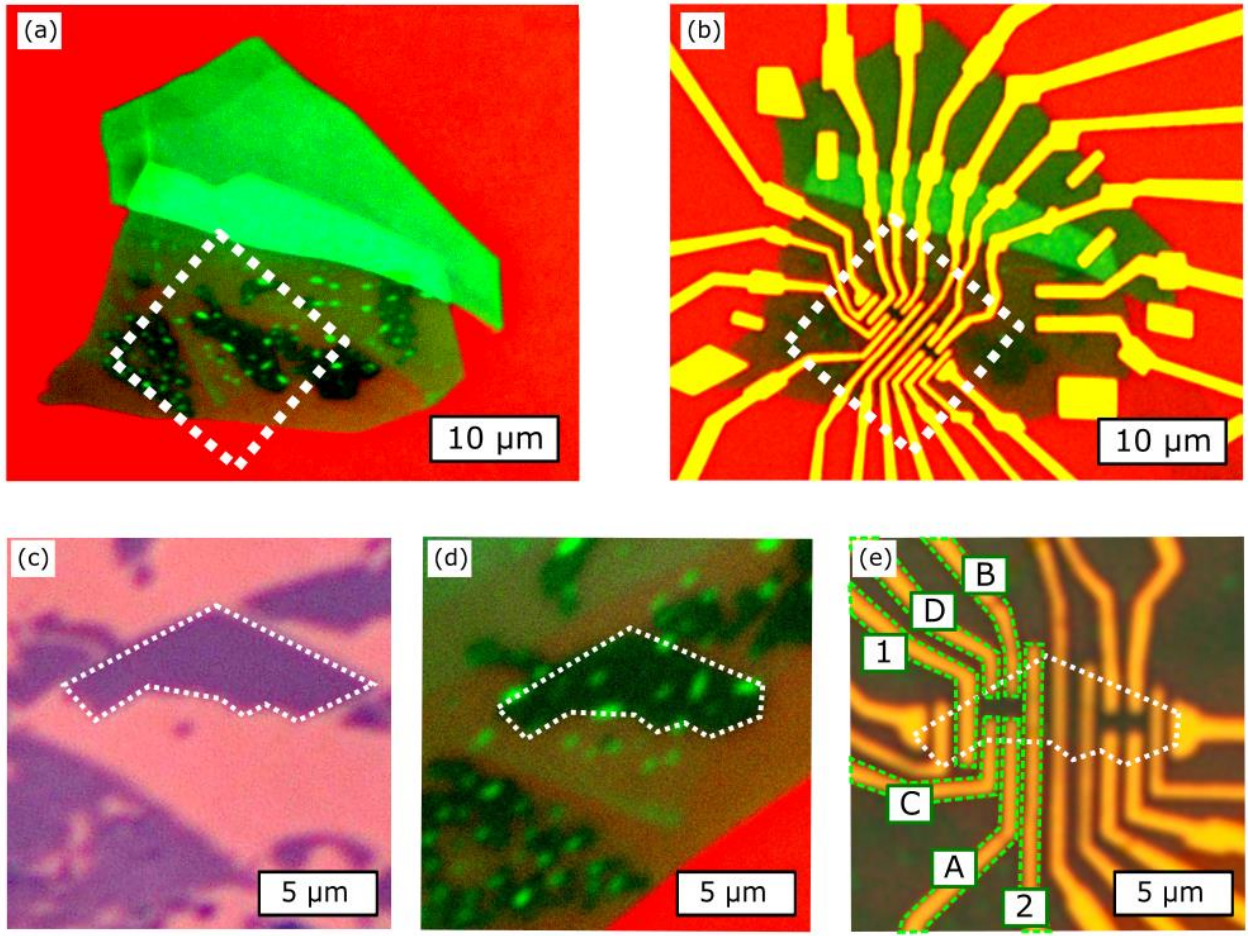

**Fig S2** Enhanced-contrast optical images of the device fabrication process (a) Fully encapsulated 1L-MoSe$_2$ crystal before the fabrication of the contacts. (b) Final device. (c-e) Zoom in of the region indicated by a dotted square in panels (a) and (b) at the different stages of the fabrication process. (c) Exfoliated 1L-MoSe$_2$ flake on SiO$_2$ before processing. (d) Same flake shown in (c) after encapsulation with top bilayer h-BN and bottom multilayer h-BN. (e) Final device with the fabricated contacts on top of the BN/MoS$_2$/BN stack. The electrodes used for the measurements of the main text are highlighted in green. The dashed white line indicates the edges of the 1L-MoSe$_2$ flake.

## S3. Electrical characterization of the 1L-MoSe$_2$ phototransistors

The DC electrical characterization of the sample is performed in the dark while keeping the sample in vacuum (10$^{-4}$ mbar). In order to obtain the electrical transport properties of the MoSe$_2$ channel, we perform four-terminal measurements in Hall-bar geometry. We apply a source-drain current on the contacts 1 and 2 (See figure 1a in the main text) and measure the voltage drop along the channel using the Hall contacts A and C. We remark that using the contacts that only partially cover the channel is preferable for the characterization of the intrinsic electrical properties of the MoSe$_2$ channel, since this allows to prevent the formation of depletion regions near the metal contacts.[4]

Figure S3a shows a transfer characteristic for the 1L-MoSe$_2$ phototransistor, presenting a clear n-type behavior. We extract the threshold gate voltage ($V_{th}$) of 19 V as the gate voltage at which the conductivity starts to increase. We estimate a field-effect mobility of about 17 cm$^2$/V.s from the linear fit to the transfer curve, for the range of gate voltage ($V_{gate} > V_{th}$) with linear dependence of conductivity. Figure S3b shows the four-terminal I-V characteristics of the phototransistor. The ohmic response of the channel can be readily observed from the linearity of the obtained

I-Vs. The inset in Figure S3b shows the square resistance of the MoSe$_2$ channel, $R_{sq}$, obtained as the slope of the linear fit to the I-V divided by the length-to-width ratio of the MoSe$_2$ channel, as a function of the gate voltage.

In our device geometry, encapsulation of the MoSe$_2$ channel with h-BN reduces the influence of the adsorbate molecules on the MoSe$_2$ surface and prevents charge scatterings due to interface impurities and the Si substrate, which largely reduces the hysteresis in the charge transport measurements. Moreover, the bilayer h-BN plays the role of a tunnel barrier for injection of charge carriers, preventing the level pinning at the contacts.

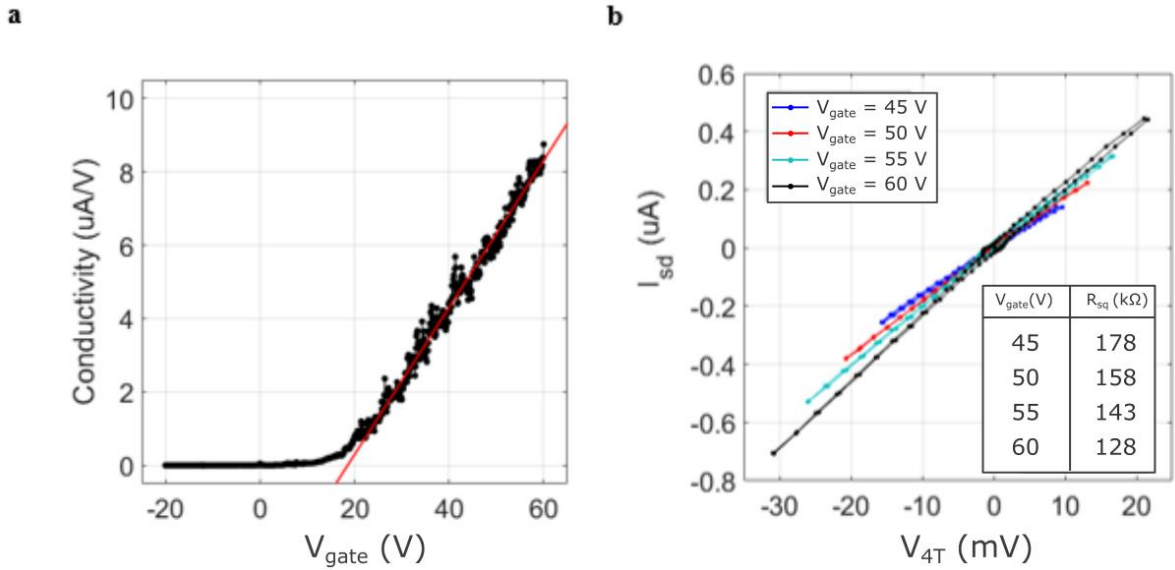

**Fig S3** Electrical characterization of the channel in 4-terminal geometry. (a) Channel conductivity as a function of gate voltage. The red line is a linear fit to the data for $V_{gate} > V_{th.}$ (b) I-V characteristics of the channel and the square resistances, estimated for the gate voltages of 40 to 60 V (shown in the legend).

## S4. Color map of the CPC amplitude as a function of $V_{ds}$ and $V_{gate}$ for illumination at normal incidence.

Figure S5 shows a colormap of the CPC amplitude $C$ as a function of the drain-source and gate voltages for $\phi = 0°$. The value of $C$ remains near zero regardless of the applied voltages. This allows us to rule out that the dominant contribution to our observed CPC signals is a Berry phase-induced CPGE, since it should become maximal for normal incidence. This measurement also rules out that our signals have a significant contribution from the valley-Hall effect, since such effect would appear as a nonzero contribution to the CPC linear with the drain-source voltage. The absence of the valley-Hall effect in our device can be understood since this effect has been reported for studies on highly n-doped devices, and it increases with the gate voltage. In our device, the 1L-MoSe$_2$ channel only starts to open for $V_{gate} > 20$ V. Thus, a much larger doping could be required for observing the valley-Hall effect.

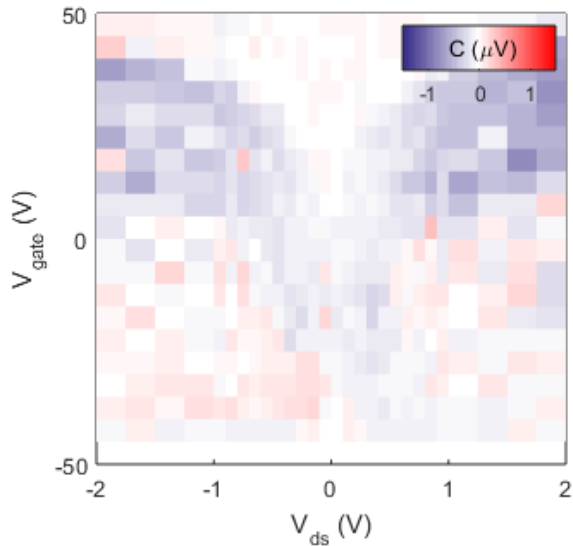

**Fig S5** Colormap of the CPC amplitude, $C$ as a function of the drain-source and gate voltages, $V_{ds}$ and $V_{gate}$ for normal incidence angle, $\phi = 0$ degrees.

## S5. Comparison of the photovoltage and photocurrent measurements and consistency checks

Figure S6 shows the helicity-dependent open-circuit photovoltage and short-circuit photocurrent measured in two sets of electrodes: [A, B] and [1, 2]. These results are representative for a wider range of checks that we performed, where we always found a linear relation between the observed values for $C$, $L_1$ and $L_2$ in the current and voltage signals. This photoresponse can thus be measured equivalently as current or voltage signals on our device. In addition, for $C$ we observed no dependence on the orientation of the linear polarization for the laser beam incident on the $\lambda/4$ plate.

Finally, Figs. 2b and 3c (main text) show that the spectral dependence of $C$ is preserved for two different sets of electrodes, further ruling out a role for specific contacts or standing-wave effects between electrodes. Our full range of consistency checks allows us to rule out that effects at specific electrodes, and effects from confining light between the micron-scale metallic electrode structure, give a significant contribution to the helicity-dependent signals that we analyze.

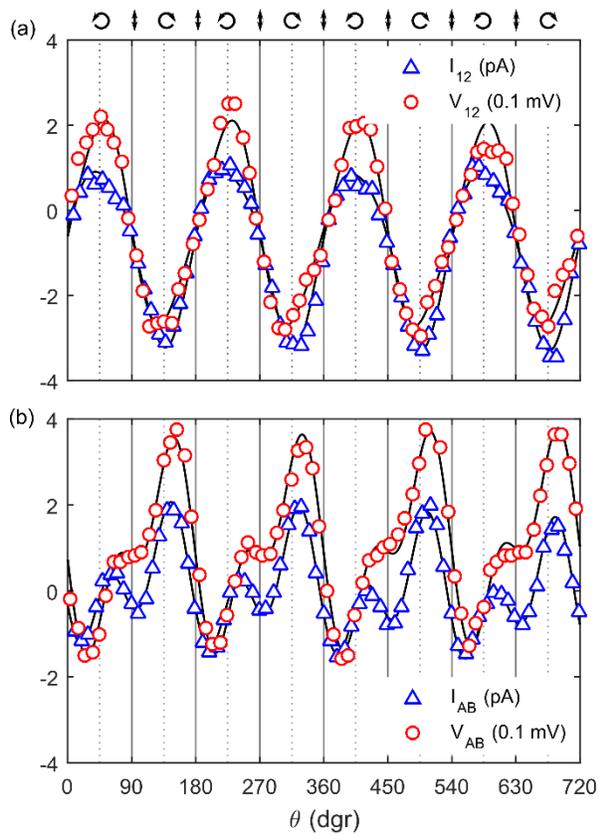

**Fig S6** Helicity-dependent open-circuit photovoltage (red circles) and short-circuit photocurrent (blue circles) for electrodes [1, 2] (a) and [A, B] (b) at $V_{gate} = 0$ V and $V_{ds} = 0$, as a function of the waveplate angle ($\theta$). The black solid lines are fittings to the phenomenological equation (1) in the main text. Except for a scale factor, the $\theta$ dependence of the photovoltage and photocurrent are very similar, as expected from the linear I-V of the semiconductor channel.

## S6. Theoretical analysis of the photogalvanic and photon drag effects

### I. General description of photogalvanic and photon drag effects

The theoretical foundations of the photogalvanic and photon drag effects (PGE and PDE) trace back to several decades ago.[5–7] More specifically, they are characterized by a DC current generated by a time-varying electric field, with amplitude proportional to the square of the applied field. This current is generated by photoelectrons which are excited by optical (vertical in the band structure) transitions and, depending on its microscopic origin, can depend on the polarization (linear PCE and PDE) or the helicity (circular PGE and PDE) of the applied field. Recent studies on derivations of PGE and PDE rely on the nonlinear susceptibility [8], Floquet theory [9] and the kinetic equation approach.

Let us consider the situation of a 2D material illuminated by a monochromatic light source, with complex electric field defined as the plane wave

$$E_j(\vec{r}, t) = E_j e^{i\omega t + i\vec{q}\cdot\vec{r}} + E_j^* e^{i\omega t - i\vec{q}\cdot\vec{r}}, \tag{1}$$

where the subindices $i$, $j$ and $k$ stand for the Cartesian coordinates, $\omega$ is the angular frequency and the wave vector $\vec{q}$ can be expressed in spherical coordinates as

$$\vec{q} = -q(\sin(\phi)\cos(\alpha), \sin(\phi)\sin(\alpha), \cos(\phi)). \tag{2}$$

We assume that the incident light forms a polar angle $\phi$ and an azimuthal angle $\alpha$ with the 2D plane (see Figure 1 in the main text). We can write down the electric field (as well as the vector potential $\vec{A} = \vec{E}/i\omega$) as

$$\vec{E} = \begin{pmatrix} E_x \\ E_y \\ E_z \end{pmatrix} = E_0 \begin{pmatrix} -i\sin(2\theta)\sin(\alpha) + (1 - i\cos(2\theta))\cos(\phi)\cos(\alpha) \\ i\sin(2\theta)\cos(\alpha) + (1 - i\cos(2\theta))\cos(\phi)\sin(\alpha) \\ -(1 - i\cos(2\theta))\sin(\phi) \end{pmatrix}, \tag{3}$$

with $E_0$ as the magnitude of the applied electric field and $\theta$ as the angle between the fast axis of the $\lambda/4$ waveplate and the initial linear polarization of the light. $\theta = \pi/4$, $(3\pi/4)$ for left (right) circularly polarized light. For future purpose, here we write $(\vec{E} \times \vec{E}^*)$ as

$$\begin{aligned}
(\vec{E} \times \vec{E}^*)_x &= -2i\cos(\alpha)\sin(2\theta)\sin(\phi) E_0^2 \\
(\vec{E} \times \vec{E}^*)_y &= -2i\sin(\alpha)\sin(2\theta)\sin(\phi) E_0^2 \\
(\vec{E} \times \vec{E}^*)_z &= -2i\sin(2\theta)\cos(\phi) E_0^2
\end{aligned} \tag{4}$$

The light-induced current density $\vec{J}$ inside the material can be generically written in series of the Cartesian components $(l, j, k)$ of the electric field $\vec{E}$. Per component of $\vec{J}$ this gives

$$J_l = \sigma_{lj} E_j e^{-i\omega t + i\vec{q}\vec{r}} + \sigma_{ljk}^{(2')} E_j E_k e^{-2i\omega t + 2i\vec{q}\vec{r}} + \sigma_{ljk}^{(2)} E_j E_k^* + \cdots, \tag{5}$$

where $\sigma_{lj}$ is a second rank tensor and $\sigma_{ljk}^{(2)}$, and $\sigma_{ljk}^{(2')}$ are third rank tensors. The first and second terms in the right correspond respectively to a linear AC current (at optical frequency) in response to the electric field and an AC current of twice the frequency of the radiation, responsible for second harmonic generation. The relevant term for us is the third term, which corresponds to a DC current $J_l^{DC}$ in response to the oscillating field:

$$J_l^{DC} = \sigma_{ljk}^{(2)} E_j E_k^*, \tag{6}$$

By doing a Taylor expansion over the wave vector $q$ we can rewrite $J_l^{DC}$ as

$$J_l^{DC} = \sigma_{ljk}^{(2)}(\omega, \vec{q})E_j E_k^* = \chi_{ljk}(\omega)E_j E_k^* + T_{ljk\mu}(\omega)q_\mu E_j E_k^* + \cdots \quad (7)$$

Here $\chi_{ljk} = \sigma_{ljk}^{(2)}(\omega, 0)$ does not depend on the radiation wave vector $\vec{q}$ and is responsible for the photogalvanic effect, $\vec{J}^{PGE}$, while $T_{ljk\mu}$ accounts for the photon drag effect $\vec{J}^{PDE}$, linear with factors $q_l$.

*Requirement of inversion symmetry breaking*

We now show that the absence of inversion symmetry is necessary for getting nonzero photogalvanic and photon drag effect. First, we note that $J_l^{DC}$ is antisymmetric (changes its sign) under inversion of the spatial coordinates $x$, $y$, $z \to -x$, $-y$, $-z$, while the object $E_j E_k^*$ is symmetric under that transformation. In consequence, if the inversion transformation is a symmetry of the studied system, $J_l^{DC}$ cannot have any dependence on $E_j E_k^*$ and $\chi_{ljk}$ must be zero. In other words, the photogalvanic effect can only emerge in systems with broken inversion symmetry.

*Linear and circular photogalvanic and photon drag effect*

Next, we observe that, since the current density must be real, it cannot change under complex conjugation. In consequence, from equation (7) we get $\chi_{lkj}^* = \chi_{ljk}$. Therefore, the real part of $\chi_{ljk}$ is symmetric under coordinate exchange while its imaginary part is antisymmetric under this operation. This allows us to rewrite the photogalvanic current as follows,

$$J_l^{PGE} = \chi_{ljk} E_\beta E_\gamma^* = \chi_{ljk}^{sym} E_j E_k^* + i\chi_{ljk}^{antisym} E_j E_k^*, \quad (8)$$

or, using the transformation under permutation of the subindices $j$ and $k$,

$$J_l^{PGE} = \frac{1}{2}\chi_{ljk}^{sym}(E_j E_k^* + E_k E_j^*) + \frac{1}{2}\chi_{ljk}^{antisym}(E_j E_k^* - E_k E_j^*) \equiv J_l^{LPGE} + J_l^{CPGE} \quad (9)$$

We can now compare $J_l^{LPGE}$ and $J_l^{CPGE}$ with the Stokes parameters:

$$S_1 = \frac{|E_x|^2 - |E_y|^2}{|E_x|^2 + |E_y|^2}; \quad S_2 = \frac{E_x E_y^* + E_x^* E_y}{|E_x|^2 + |E_y|^2}; \quad S_3 = i\frac{E_x E_y^* - E_x^* E_y}{|E_x|^2 + |E_y|^2} \quad (10)$$

We see that $J_l^{LPGE}$ is proportional to $S_2$, which accounts for the linearly polarized radiation (linear photogalvanic effect), while $J_l^{CPGE}$ is proportional to $S_3$, and, therefore, it is sensitive to the circularly polarized radiation (circular photogalvanic effect).

Finally, the totally antisymmetric Levi-Civita tensor $\epsilon_{sjk}$ can be used to contract $\chi_{ljk}^{antisym}$ to only one pseudo vector index,

$$\sum_{jk}\chi_{ljk}^{antisym}(E_j E_k^* - E_k E_j^*) = i\sum_{sjk}2\gamma_{ls}\epsilon_{sjk}(E_j E_k^* - E_k E_j^*) = i\sum_s \gamma_{ls}(\vec{E} \times \vec{E}^*)_s, \quad (11)$$

where $\gamma_{ls}$ is a second rank pseudo-tensor and $l$ and $s$ stand for Cartesian coordinates. Thus $J_l^{CPGE}$ can be expressed as [5]

$$J_l^{\text{CPGE}} = i \sum_j \gamma_{lj}(\vec{E} \times \vec{E}^*)_j \ , \tag{12}$$

It is convenient to separate $J_l^{\text{PDE}}$ in a similar fashion into its circular and linear polarization sensitive components, $J_l^{\text{CPDE}}$ and $J_l^{\text{LPDE}}$. For $J_l^{\text{CPDE}}$ we get:

$$J_l^{\text{CPDE}} = i \sum_{jk} T_{ljk} q_j (\vec{E} \times \vec{E}^*)_k \ , \tag{13}$$

At this point it is worth noting that we have not still made any assumption about the physical origin of $J_l^{CPGE}$. Thus, equations (12) and (13) are completely general and must hold regardless of the underlying physical mechanism.

### II. Symmetry arguments for the CPC in monolayer TMDCs.

In the following, we use symmetry arguments to determine the nonzero components of the tensors $\gamma_{ij}$ and $T_{lsj}$, as defined in equations (12) and (13). This allows to extract constraints for the dependence of $J_l^{\text{CPGE}}$ and $J_l^{\text{CPDE}}$ on the illumination angles $\alpha$ and $\phi$. We remark again that, since equations (12) and (13) must hold regardless of the physical origin of the CPC, the discussion below is completely general.

In order of decreasing symmetry we analyze three cases: $D_{3h}$, $C_{3v}$, and *Single mirror-plane* symmetry. We find that crystal structures belonging to the high-symmetry class $D_{3h}$ cannot support any CPGE. For the case of $C_{3v}$ symmetry with an oblique incidence angle $\phi$, only $\gamma_{xy}$ can have a nonzero value, which then gives a nonzero CPGE, but always with the property that it flips signs upon reversal of $\phi$ (in conflict with a BC origin). Systems with only one mirror symmetry can not only have nonzero $\gamma_{xy}$ but also nonzero $\gamma_{yz}$ and $\gamma_{xz}$, allowing for BC-CPGE. We conclude that our experimental results for low source-drain voltage are only compatible with, at most, one mirror-plane symmetry, since otherwise $\gamma_{xz}$ and $\gamma_{yz}$ cancel out and, therefore, the photocurrent cannot be preserved upon inversion of the incidence angle, $\phi$.

## $D_{3h}$ symmetry

The scenario of a 1L-TMDC system with ideal mirror symmetry with respect to the crystal plane (symmetric environments and external fields) give the system the high $D_{3h}$ symmetry. This has a three-fold rotation symmetry around the $z$ axis, defined by the operator $C_3$, three two-fold axes perpendicular to $C_3$, a mirror plane in the $xy$ plane, defined by $\sigma_h = \begin{pmatrix} 1 & 0 & 0 \\ 0 & 1 & 0 \\ 0 & 0 & -1 \end{pmatrix}$ and three mirror vertical planes with respect to the $xy$ plane $\sigma_v$. The improper rotation is $\sigma_h C_3$.

In the section directly below here on $C_{3v}$ symmetry we derive that the CPGE current is

$$\vec{j}^{CPGE} = i \begin{pmatrix} 0 & \gamma_{xy} & 0 \\ -\gamma_{xy} & 0 & 0 \\ 0 & 0 & 0 \end{pmatrix} \begin{pmatrix} (\vec{E} \times \vec{E}^*)_x \\ (\vec{E} \times \vec{E}^*)_y \\ (\vec{E} \times \vec{E}^*)_z \end{pmatrix} = i \begin{pmatrix} \gamma_{xy}(\vec{E} \times \vec{E}^*)_y \\ -\gamma_{xy}(\vec{E} \times \vec{E}^*)_x \\ 0 \end{pmatrix}. \tag{14}$$

This result for $C_{3v}$ can be extended to the case for $D_{3h}$ by adding the requirement for the additional mirror symmetry $\sigma_h$. This brings that a pseudo-vector $(\vec{E} \times \vec{E}^*)$ becomes $-\sigma_h(\vec{E} \times \vec{E}^*)$, and gives the condition $\gamma_{xy} = 0$. Consequently, all CPGE current contributions cancels out for the $D_{3h}$ symmetry.

For CPDE, the $C_{3v}$ symmetry (below) yields

$$\vec{j}^{CPDE} = i(T_{yxz} + T_{yzx})q_z \begin{pmatrix} -(\vec{E} \times \vec{E}^*)_y \\ (\vec{E} \times \vec{E}^*)_x \\ 0 \end{pmatrix} \propto \begin{pmatrix} -\sin(\alpha)\sin(2\theta)\sin(2\phi) \\ \cos(\alpha)\sin(2\theta)\sin(2\phi) \\ 0 \end{pmatrix}. \tag{15}$$

A direct extension of this analysis shows that the additional mirror plane $\sigma_h$ does not impose further constrains on $\vec{j}^{CPDE}$. Thus, CPDE photocurrents can be present for this symmetry.

## $C_{3v}$ symmetry

If we assume a 1L-TMDC crystal symmetry in the plane, but drop the assumption of mirror symmetry with respect to the crystal plane (relevant, for example, for a 1L-TMDC sustained on a substrate), the system has $C_{3v}$ symmetry. This corresponds to a three-fold rotation symmetry around the $z$ axis $C_3$, and three mirror planes perpendicular to the $xy$ plane. The CPGE photocurrent is given by

$$j_i^{CPGE} = i\gamma_{ij}(\vec{E} \times \vec{E}^*)_j. \tag{16}$$

Under a $2\pi/3$ rotation

$$R = \begin{pmatrix} \cos(2\pi/3) & \sin(2\pi/3) & 0 \\ -\sin(2\pi/3) & \cos(2\pi/3) & 0 \\ 0 & 0 & 1 \end{pmatrix},$$

$\vec{E} \times \vec{E}^*$ becomes $R(\vec{E} \times \vec{E}^*)$ and $\vec{j}^{CPGE}$ becomes $R\vec{j}^{CPGE}$. Since $\gamma$ should remain the same under the rotational symmetry, we have

$$R\vec{j}^{CPGE} = i\gamma R(\vec{E} \times \vec{E}^*). \tag{17}$$

By replacing (26) into (27) we obtain

$$R\gamma = \gamma R. \tag{18}$$

An additional constraint is given by the mirror symmetry. There are three mirror planes perpendicular to the $xy$ plane, and we assume that the angle between the mirror plane and the $x$ axis is $\psi$. Under the mirror reflection, characterized by the operator $= \begin{pmatrix} \cos(2\psi) & \sin(2\psi) & 0 \\ \sin(2\psi) & -\cos(2\psi) & 0 \\ 0 & 0 & 1 \end{pmatrix}$, $\gamma$ should remain the same, while $\vec{j}^{CPGE}$ becomes $M\vec{j}^{CPGE}$. As a pseudo-vector, $(\vec{E} \times \vec{E}^*)$ becomes $-M(\vec{E} \times \vec{E}^*)$. Therefore, we obtain

$$M\gamma = -\gamma M. \tag{19}$$

Combining the constraints from rotational and mirror symmetries, we conclude that $\gamma_{ij}$ has only one independent parameter:

$$\gamma_{ij} = \begin{pmatrix} 0 & \gamma_{xy} & 0 \\ -\gamma_{xy} & 0 & 0 \\ 0 & 0 & 0 \end{pmatrix}, \tag{20}$$

and the CPGE current must have a form of

$$\vec{j}^{CPGE} = i\begin{pmatrix} 0 & \gamma_{xy} & 0 \\ -\gamma_{xy} & 0 & 0 \\ 0 & 0 & 0 \end{pmatrix}\begin{pmatrix} (\vec{E} \times \vec{E}^*)_x \\ (\vec{E} \times \vec{E}^*)_y \\ (\vec{E} \times \vec{E}^*)_z \end{pmatrix} \propto \gamma_{xy}\begin{pmatrix} \sin(\alpha)\sin(2\theta)\sin(\phi) \\ -\cos(\alpha)\sin(2\theta)\sin(\phi) \\ 0 \end{pmatrix}. \tag{21}$$

This description yields a nonzero CPGE current that indeed shows a $\sin(2\theta)$ dependence in the polarization control. Further, it only yields nonzero CPGE currents for nonzero angles $\phi$. For this symmetry group the CPGE current changes sign when the incidence angle is switched from $\phi$ to $-\phi$.

For the CPDE, we have $J_l^{CPDE} = iT_{ljk}q_j(\vec{E} \times \vec{E}^*)_k$. Imposing that, for a transverse electromagnetic wave, the vector $(\vec{E} \times \vec{E}^*)$ should be along the same direction as the photon momentum $\vec{q}$, we get $q_j(\vec{E} \times \vec{E}^*)_k = q_k(\vec{E} \times \vec{E}^*)_j$. Imposing the invariance of $\vec{j}^{CPDE}$ under a rotation of $2\pi/3$, $\vec{j}^{CPDE'} = R\vec{j}^{CPDE}$ we get

$$T_{zij}R_{ik}q_k R_{jl}(\vec{E} \times \vec{E}^*)_l = T_{zij}q_i(\vec{E} \times \vec{E}^*)_j. \tag{22}$$

From each $q_i(\vec{E} \times \vec{E}^*)_j$ we get

$$T_{zxz} + T_{zzx} = T_{zyz} + T_{zzy} = T_{zxy} + T_{zyx} = 0;$$
$$T_{zyy} = T_{zxx}. \tag{23}$$

Also, from $j_x$ and $j_y$ we get

$$T_{xzz} = 0; T_{xyy} = -T_{xxx}; T_{yxx} = \frac{\sqrt{3}}{2}T_{xxx}; T_{yxy} + T_{yyx} = -2T_{xxx};$$
$$T_{yxz} + T_{yzx} = -(T_{xyz} + T_{xzy}); T_{yyz} + T_{yzy} = T_{xxz} + T_{xzx}; T_{yyy} = -\frac{1}{2}(T_{xxy} + T_{xyx}). \tag{24}$$

Combining this with the restrictions imposed by the mirror symmetry we find that only $T_{yxz} + T_{yzx} = -(T_{xyz} + T_{xzy}) \equiv \chi$ remains and

$$\vec{j}^{CPDE} = i(T_{yxz} + T_{yzx})q_z \begin{pmatrix} -(\vec{E} \times \vec{E}^*)_y \\ (\vec{E} \times \vec{E}^*)_x \\ 0 \end{pmatrix} \propto \begin{pmatrix} -\sin(\alpha)\sin(2\theta)\sin(2\phi) \\ \cos(\alpha)\sin(2\theta)\sin(2\phi) \\ 0 \end{pmatrix}. \tag{25}$$

Remarkably, under $C_{3v}$ symmetry we find that a CPGE must have a $\sin(\phi)$ dependence, while a CPDE must have a $\sin(2\phi)$ dependence. They can thus be distinguished by their dependence on $\phi$.

*Single mirror-plane symmetry*

Now we consider a case of even lower symmetry: that of a 1L-TMDC system that has just one mirror plane which is perpendicular to the $xy$ plane. In a real device, the presence of asymmetric electrodes and strain gradients is expected to lead to this low-symmetry situation (or even lower symmetry). In particular, this scenario is relevant for strained monolayer MoSe$_2$, since the lowest stiffness for deformation occurs along the armchair direction in the crystal. We cannot assume that this crystal direction has a known relation with the $xy$ coordinate frame (defined by the experimental geometry with electrodes, see Fig. 1 main text). We therefore introduce the angle $\psi$ to describe angle between the mirror plane and the $x$ axis.

Under the mirror reflection, $\vec{j}^{CPGE}$ becomes $M\vec{j}^{CPGE}$, where $M = \begin{pmatrix} \cos(2\psi) & \sin(2\psi) & 0 \\ \sin(2\psi) & -\cos(2\psi) & 0 \\ 0 & 0 & 1 \end{pmatrix}$. Once more, as a pseudovector, $(\vec{E} \times \vec{E}^*)$ becomes $-M(\vec{E} \times \vec{E}^*)$. Therefore, we obtain $M\gamma = -\gamma M$. The absence of the $2\pi/3$ rotational symmetry in this case allows more independent parameters to appear in $\gamma_{ij}$ and the CPGE current takes the form

$$\vec{j}^{CPGE} = i\begin{pmatrix} 0 & \gamma_{xy} & \gamma_{xz} \\ -\gamma_{xy} & 0 & \gamma_{yz} \\ \gamma_{zx} & \gamma_{zy} & 0 \end{pmatrix} \begin{pmatrix} (\vec{E} \times \vec{E}^*)_x \\ (\vec{E} \times \vec{E}^*)_y \\ (\vec{E} \times \vec{E}^*)_z \end{pmatrix} \propto \begin{pmatrix} [\gamma_{xy}(\sin(\alpha)\sin(\phi) + \gamma_{xz}\cos(\phi)]\sin(2\theta) \\ [-\gamma_{xy}(\cos(\alpha)\sin(\phi) + \gamma_{yz}\cos(\phi)]\sin(2\theta) \\ 0 \end{pmatrix}, \tag{26}$$

where we assumed that the $j_z^{CPGE}$ component must be zero for a single-layer crystal.

The mirror symmetry yields the additional conditions:

$$\gamma_{xz} = \frac{1 - \cos(2\psi)}{\sin(2\psi)} \gamma_{yz} \quad \text{and} \tag{27}$$

$$\gamma_{zx} = \frac{1 - \cos(2\psi)}{\sin(2\psi)} \gamma_{zy}. \tag{28}$$

Thus, we get a CPGE contribution independent of $\alpha$ and changing as $\cos(\phi)$. In the next section we will show that this angular dependence is required for a Berry curvature-induced CPGE.

Now we consider the CPDE. For simplicity, we assume that the mirror symmetry is from $x$ to $-x$. We get

$$j_x^{CPDE} = 2\sin(2\theta)\left\{T_{xzz}\cos^2(\phi) + \frac{T_{xyz} + T_{xzy}}{2}\sin(\alpha)\sin(2\phi) \right.$$
$$\left. + (T_{xxx}\cos^2(\alpha) + T_{xyy}\sin^2(\phi))\sin^2(\phi)\right\} ;$$
$$j_y^{CPDE} = \sin(2\theta)\left\{(T_{yxz} + T_{yzx})\cos(\alpha)\sin(2\phi) + (T_{yxy} + T_{yyx})\sin(2\alpha)\sin^2(\phi)\right\} ;$$
$$j_z^{CPDE} = \sin(2\theta)\left\{(T_{zxz} + T_{zzx})\cos(\alpha)\sin(2\phi) + (T_{zxy} + T_{zyx})\sin(2\alpha)\sin^2(\phi)\right\} ; \tag{29}$$

Here, we find that CPDE can have a contribution dependent on $\sin(2\alpha)\sin^2(\phi)$, which matches the observed angular dependence for the low-$V_{ds}$ regime.

As discussed in the main text, this symmetry analysis does not bring forward the $\sin(3\alpha)$ dependence experimentally observed for the large-$V_{ds}$ regime (although this could still come forward from asymmetric transport properties of the zig-zag and armchair directions of the MoSe$_2$ crystal).

In the particular case where an additional mirror symmetry in the $z$ direction is allowed (original $D_{3h}$ case subject to e.g. uniaxial strain), the CPGE and CPDE current is further constrained, yielding

$$\vec{J}^{CPGE} = i \begin{pmatrix} 0 & 0 & \gamma_{xz} \\ 0 & 0 & \gamma_{yz} \\ \gamma_{zx} & \gamma_{zy} & 0 \end{pmatrix} \begin{pmatrix} (\vec{E} \times \vec{E}^*)_x \\ (\vec{E} \times \vec{E}^*)_y \\ (\vec{E} \times \vec{E}^*)_z \end{pmatrix} \propto \begin{pmatrix} \gamma_{xz} \cos(\phi) \sin(2\theta) \\ \gamma_{yz} \cos(\phi) \sin(2\theta) \\ 0 \end{pmatrix}, \quad (30)$$

$$\vec{J}^{CPDE} = \begin{pmatrix} (T_{xyz} + T_{xzy}) \sin(\alpha) \sin(2\phi) \sin(2\theta) \\ (T_{yxz} + T_{yzx}) \cos(\alpha) \sin(2\phi) \sin(2\theta) \\ (T_{zyz} + T_{zzx}) \cos(\alpha) \sin(2\phi) \sin(2\theta) \end{pmatrix}, \quad (31)$$

and only the $\cos(\phi)$-dependent CPGE contribution can still appear. Also, for the CPDE current, the terms on $\sin(2\alpha)\sin^2(\phi)$ fade out. Thus, we see that our experimental results require a broken out-of-plane mirror symmetry.

### III. Berry curvature and circular photogalvanic effect

In this section, we show how a circular photogalvanic current can emerge as a consequence of a nonzero Berry curvature. We start the derivation of the photocurrent equation based on the assumption that the momentum of light is small and can be ignored. Thus, we consider only vertical optical interband transition between the initial and final bands. The photocurrent $\vec{J}$ can be derived based on the Fermi-Golden rule [10]:

$$J_i = -\frac{2\pi e\tau}{\hbar} \sum_{I,F} \frac{d^2k}{(2\pi)^2} f_{IF}(\vec{k})(v_F^i - v_I^i)\delta(\Delta E_{FI} - \omega)|D|^2 \quad (32)$$

where $v$ is the group velocity of the electron state, $\omega$ is the excitation energy, $\Delta E_{FI} = E_F(\vec{k}) - E_I(\vec{k})$ and $f_{IF}(\vec{k}) \equiv f_I(\vec{k}) - f_F(\vec{k})$ are the differences of energy and equilibrium Fermi distribution function between the initial and final states. $|I\rangle$ and $|F\rangle$ are the Bloch wavefunction of the initial and final states. $D$ is the optical transition dipole defined as

$$D = \frac{e}{m_e} \langle F| \vec{A} \cdot \vec{p} |I\rangle, \quad (33)$$

where $e$ and $m_e$ are the charge and mass of a bare electron, $\vec{A}$ is the vector potential of light and $\vec{p}$ is the momentum operator defined as $\vec{p} = (m_e/i\hbar)[\vec{r}, H]$. Here we assume $\tau$ is the relaxation time for all bands.

Now we consider the light as

$$E = \begin{pmatrix} E_x \\ E_y \\ E_z \end{pmatrix} = \begin{pmatrix} -i\sin(2\theta)\sin(\alpha) + (1 - i\cos(2\theta))\cos(\phi)\cos(\alpha) \\ i\sin(2\theta)\sin(\alpha) + (1 - i\cos(2\theta))\cos(\phi)\cos(\alpha) \\ -(1 - i\cos(2\theta))\sin(\phi) \end{pmatrix}, \quad (34)$$

with $E_0$ as the magnitude of the electric field and $\vec{A} = \vec{E}/i\omega$ Therefore under this polarized light $|D^2|$ is given as

$$|D|^2 = \left(\frac{\omega e}{m_e}\right)^2 |\langle F|E_x p_x + E_y p_y + E_z p_z|I\rangle|^2. \tag{35}$$

We further rewrite $E_a E_b^*$ (the subindices refer to Cartesian coordinates) as

$$E_a E_b^* = \{E_a E_b^*\} + [E_a E_b^*], \tag{36}$$

where $\epsilon_{sab}[E_a E_b^*] = \frac{1}{2}(\vec{E} \times \vec{E}^*)_s$ with

$$(\vec{E} \times \vec{E}^*)_x = -2i \cos(\alpha) \sin(2\theta) \sin(\phi) E_0^2 ; \tag{37}$$

$$(\vec{E} \times \vec{E}^*)_y = -2i \sin(\alpha) \sin(2\theta) \sin(\phi) E_0^2 ; \tag{38}$$

$$(\vec{E} \times \vec{E}^*)_z = -2i \sin(2\theta) \cos(\phi) E_0^2 . \tag{39}$$

In the following, we drop $\{E_a E_b^*\}$ and $|E_a|^2$ terms which do not contribute to the circular photogalvanic effect, characterized by the $\sin(2\theta)$ dependence. Therefore, we obtain

$$|D|^2 = \frac{1}{2}\left(\frac{\omega e}{m_e}\right)^2 \Big[(\vec{E}\times\vec{E}^*)_z(\langle F|p_x|I\rangle\langle I|p_y|F\rangle - \langle F|p_y|I\rangle\langle I|p_x|F\rangle) \\ -(\vec{E}\times\vec{E}^*)_y(\langle F|p_x|I\rangle\langle I|p_z|F\rangle - \langle F|p_z|I\rangle\langle I|p_x|F\rangle) \\ +(\vec{E}\times\vec{E}^*)_x(\langle F|p_y|I\rangle\langle I|p_z|F\rangle - \langle F|p_z|I\rangle\langle I|p_y|F\rangle)\Big]. \tag{40}$$

Here, the $\sin(2\theta)$ factor that governs the currents changes sign when the helicity of light is inverted.

In a 2D crystal, we can use Peierls substitution, $p_i = \frac{m_e}{\hbar}[z, \hat{H}]$ for $i = x, y$. Because translational symmetry is broken along the $z$ direction for a 2D crystal, we use $p_z = im_e/\hbar[z, \hat{H}]$ in $|D|^2$,

$$|D|^2 = \left(\frac{\omega E_0 e}{\hbar}\right)^2 \Bigg[-i\cos(\phi)\sin(2\theta)\left(\langle F|\frac{\partial \hat{H}}{\partial k_x}|I\rangle\langle I|\frac{\partial \hat{H}}{\partial k_y}|F\rangle - \langle F|\frac{\partial \hat{H}}{\partial k_y}|I\rangle\langle I|\frac{\partial \hat{H}}{\partial k_x}|F\rangle\right) \\ + \sin(\alpha)\sin(\phi)\sin(2\theta)\left(\langle F|\frac{\partial \hat{H}}{\partial k_x}|I\rangle\langle I|[z,\hat{H}]|F\rangle - \langle F|[z,\hat{H}]|I\rangle\langle I|\frac{\partial \hat{H}}{\partial k_x}|F\rangle\right) \\ -\cos(\alpha)\sin(\phi)\sin(2\theta)\left(\langle F|\frac{\partial \hat{H}}{\partial k_y}|I\rangle\langle I|[z,\hat{H}]|F\rangle - \langle F|[z,\hat{H}]|I\rangle\langle I|\frac{\partial \hat{H}}{\partial k_y}|F\rangle\right)\Bigg]. \tag{41}$$

The first term in $|D|^2$ can be related to Berry curvature (BC) for the electronic Bloch states of the $n^{th}$ band:

$$\Omega_n^z(\vec{k}) = i\hat{z} \cdot (\nabla_{\vec{k}} u_{n\vec{k}}^*) \times (\nabla_{\vec{k}} u_{n\vec{k}}) = -2\sum_{n\neq n'} \frac{\text{Im}\left(\langle u_{n\vec{k}}|\frac{\partial \hat{H}}{\partial k_x}|u_{n'\vec{k}}\rangle\langle u_{n'\vec{k}}|\frac{\partial \hat{H}}{\partial k_y}|u_{n\vec{k}}\rangle\right)}{[E_n(\vec{k}) - E_{n'}(\vec{k})]^2}. \tag{42}$$

In a 2D crystal the Berry curvature has only a nonzero component, perpendicular to the $xy$ plane (the Berry curvature behaves as a pseudoscalar). In a N-band system, the BC of the $n^{th}$ band comes from all the other N - 1 bands.

For a simple two-band approximation, $F$ stands for the conduction band (CB) and $I$ for the valance band (VB) with the definition of Berry curvature,

$$\Omega_F^z(\vec{k}) = -\Omega_I^z(\vec{k}) = \frac{2 \, \text{Im} \left( \langle CB | \frac{\partial \hat{H}}{\partial k_x} | VB \rangle \langle VB | \frac{\partial \hat{H}}{\partial k_y} | CB \rangle \right)}{\left[ E_{CB}(\vec{k}) - E_{VB}(\vec{k}) \right]^2}. \tag{43}$$

This approximation allows us to simplify $|D|^2$ as

$$|D|^2 = \left( \frac{\Omega E_0 e}{\hbar} \right)^2 \Big[ -i \cos(\phi) \sin(2\theta) \, \Omega_F^z(\vec{k}) (\Delta E_{FI})^2$$
$$+ \sin(\alpha) \sin(\phi) \sin(2\theta) \left( \langle F | \frac{\partial \hat{H}}{\partial k_x} | I \rangle \langle I | z | F \rangle - \langle F | z | I \rangle \langle I | \frac{\partial \hat{H}}{\partial k_x} | F \rangle \right) \Delta E_{FI} \tag{44}$$
$$- \cos(\alpha) \sin(\phi) \sin(2\theta) \left( \langle F | \frac{\partial \hat{H}}{\partial k_y} | I \rangle \langle I | z | F \rangle - \langle F | z | I \rangle \langle I | \frac{\partial \hat{H}}{\partial k_y} | F \rangle \right) \Delta E_{FI} \Big].$$

The first term of $|D^2|$ is from Berry curvature $\Omega_F^z(\vec{k})$ and shows that this contribution to the CPGE is independent of $\alpha$, and maximal for normal incidence, $\phi = 0$. As discussed in the main text and in section S4, we do not find any contribution to CPC that satisfies this angular dependence. It is worth noting that, from the general definition of $J_1^{CPGE}$, equation (12), we find that a CPGE contribution changing as $\cos(\phi) \sin(2\theta)$ is associated with the matrix elements $\gamma_{xz}$ and $\gamma_{yz}$. The symmetry arguments discussed above confirm that these matrix elements can only be nonzero if the device symmetry is reduced to, at most, a single mirror plane. Therefore, the $D_{3h}$ symmetry of 1L-MoSe$_2$ must be reduced (for example from device asymmetries or strain gradients) in order to allow for a Berry curvature-induced CPGE (BC-CPGE).